\documentclass[twocolumn,showpacs,pra,aps,superscriptaddress,floatfix]{revtex4}
\usepackage{graphicx}
\begin{document}
\def\la{\langle}
\def\ra{\rangle}
\def\om{\omega}
\def\o0{\omega_0}
\def\Om{\Omega}
\def\vep{\varepsilon}
\def\wh{\widehat}
\def\P0{\wh{\cal P}_0}
\def\dt{\delta t}
\newcommand{\beq}{\begin{equation}}
\newcommand{\eeq}{\end{equation}}
\newcommand{\beqa}{\begin{eqnarray}}
\newcommand{\eeqa}{\end{eqnarray}}
\newcommand{\intf}{\int_{-\infty}^\infty}
\newcommand{\into}{\int_0^\infty}

\title{Time scale of forerunners in quantum tunneling}

\author{G. Garc\'{\i}a Calder\'on}
\email{gaston@fisica.unam.mx}
\affiliation{Instituto de F\'{\i}sica,
Universidad Nacional Aut\'onoma de M\'exico\\
Apartado Postal {20 364}, 01000 M\'exico, D.F., M\'exico}

\author{Jorge Villavicencio}
\email{villavics@uabc.mx}
\affiliation{Instituto de F\'{\i}sica,
Universidad Nacional Aut\'onoma de M\'exico\\
Apartado Postal {20 364}, 01000 M\'exico, D.F., M\'exico}
\affiliation{Facultad de Ciencias,
Universidad Aut\'onoma de Baja California\\
Apartado Postal 1880, 22800 Ensenada, Baja California, M\'exico}

\author {F. Delgado}
\email{qfbdeacf@lg.ehu.es}
\affiliation{Departamento de Qu\'{\i}mica-F\'{\i}sica,
UPV-EHU,\\
Apartado 644, 48080 Bilbao, Spain}

\author {J. G. Muga}
\email{qfpmufrj@lg.ehu.es}
\affiliation{Departamento de Qu\'{\i}mica-F\'{\i}sica,
UPV-EHU,\\
Apartado 644, 48080 Bilbao, Spain}


\begin{abstract}
The forerunners preceding the main tunneling signal of the wave created 
by a source with a sharp onset or by a quantum shutter, have been 
generally associated with over-the-barrier (non-tunneling) components. 
We demonstrate that, while this association is true for distances which are 
larger than the penetration lenght,  
for smaller distances the forerunner is dominated by under-the-barrier 
components. We find that its characteristic arrival time is inversely 
proportional to the difference between the barrier energy and the 
incidence energy, a tunneling time scale different from both the phase 
time and the B\"uttiker-Landauer (BL) time.
\end{abstract}


\maketitle

\section{INTRODUCTION}

Tunneling is one of the paradigms of quantum theory.
Textbooks only discuss its stationary aspects,   
emphasizing the non-vanishing transmission
probability for monochromatic wavefunctions with energies below the barrier
maximum. In recent times,  
since the seminal work of B\"uttiker 
and Landauer \cite{BL82}, its time dependent aspects have been also
investigated to find characteristic time scales that summarize,
together with the transmittance,
the system behavior. Much of this work has been controversial, as several
authors have proposed and defended different
``tunneling times''.
In fact each has its own virtues, weaknesses, and 
range of applicability. For a recent and rather extensive  
multi-author review see \cite{MSE02};
for previous reviews see \cite{HS89,LA90,LM94,Ghose99}.
An analysis of the involved 
non-commuting observables (the projectors that
determine the final transmission and
the probability to find the particle in the barrier region)
shows that, from a fundamental perspective, 
there is no unique tunneling time, because several quantizations are possible
due to different
operator orderings and defining criteria \cite{BSM94}.  

This does not mean though that the timing question should be abandoned.
Rather, it is necessary to specify more precisely how to time the quantum
particle in the tunneling regime. Different specifications 
lead to different relevant time scales; in other words, the importance
of the differently defined times rests on the context where they
become physically significant quantities.
For example, the traversal time of B\"uttiker and Landauer (BL time)\cite{BL82}, $\tau=L/v_{sc}$, given by the barrier length
$L$ divided by the ``semiclassical'' velocity $v_{sc}=[2(V-E)/m]^{1/2}$,
marks the transition from sudden to adiabatic regimes
for an oscillating barrier \cite{BL82}, and  determines
the rotation of the spin
in a weak magnetic field in opaque conditions \cite{Buttiker83};
whereas the average over wavepacket components of the (monochromatic)
``phase times'' provides the mean arrival time of the transmitted wave
packet \cite{BSM94,ML00}. 
These two time scales may be very different.
The B\"uttiker-Landauer time increases with  
decreasing energies up to a finite value,
whereas $\tau^{Ph}(0,L)$ (the so called
extrapolated phase time) diverges as $E\to 0$, and   
tends for increasing $L$ to a constant value, 
$2\hbar/[v_{sc}(2mE)^{1/2}]$. 
This later property implies that the arrival of the transmitted wave 
becomes independent of $L$ (Hartman effect \cite{Hartman62}), although  
the independence only holds  
until a certain critical length $L_c$ \cite{BSM94}
where above-the-barrier components start to dominate. For 
$L>L_c$ the mean arrival time depends on $L$ linearly.
While $\tau$ and $\tau^{Ph}$ are surely the most frequently
invoked tunneling times, they do not exhaust all 
timing questions.

In particular, in Ref.\ \cite{MB00}, the time $t_{tr}=t_{tr}(x)$
that characterizes the transition  
from the transient to the stationary regime for a wave formed by
a point source with a sharp onset, was identified and
the time of arrival, $t_p=t_p(x)$,  of the peak
of the forerunner at a point $x$ in opaque conditions
was also identified. Here ``opaque'' means  
$x\kappa_0 \gg 1$, where $\kappa_0=[2m(V-E_0)]^{1/2}/\hbar$, 
and $E_0=\hbar \o0$.
This particular time scale turned out to be (surprisingly)
proportional to the BL time $\tau$ {\it i.e.},
\beq
\label{tf}
t_p=\tau/3^{1/2},
\eeq
even though the time-frequency analysis of the forerunner showed
that it was composed by frequencies above
threshold. In other words, it corresponded to non-tunneling.
In spite of its frequency content, the forerunner's peak 
``travels'' with a velocity proportional to $v_{sc}$, $v_p=3^{1/2}v_{sc}$, 
which increases with decreasing energies, and its intensity 
diminishes exponentially as it progresses along the coordinate $x$.   
Other works had already pointed out the dominance of non-tunneling 
components in the forerunner \cite{RMFP90,RMA91,TKF87,APJ89,BM96} 
but had not characterized its time dependence. Their main objective
was to show that a previous prediction by Stevens \cite{Stevens}
(who believed that a tunneling monochromatic
$\om_0$-front, associated with a pole at $\om_0$ in the complex
frequency plane would arrive at $\tau$) 
did not hold because of the effect 
of the saddle point or other critical points, such as resonances 
in a square barrier model \cite{BM96}. 
(For the source-with-a-sharp-onset model
the frequency of the saddle is given by $\omega_s=E_s/\hbar$,
where the saddle energy, $E_s=V+x^2m/2t^2$, coincides with 
the energy of a classical particle traveling from the source to
$x$ in a time $t$.) B\"uttiker and Thomas \cite{BT} proposed to enhance
the importance of the monochromatic front associated with $\omega_0$ 
compared to the forerunners by limiting the frequency band of the 
source or of the detector; it was shown later in \cite{MB00}
that the monochromatic front could not be seen in 
opaque conditions even with the frequency band limitation.     
However, a clear separation of the amplitude into two terms, one 
associated with saddle and forerunner and the other with the pole 
and the ``monochromatic front'', is only possible for opaque conditions.     
Non semiclassical conditions have been much less investigated \cite{Nimtz}, 
even though these are actually easier to observe because of 
the stronger signal. 
In this paper we shall be mainly concerned with them.   

The motivation of the present paper is 
a recent publication of two of us where the 
time evolution of an initial cutoff wave truncated at the
left edge of a square barrier (shutter problem) \cite{gcv01} was examined. 
There it was found that the probability density at the barrier edge $x=L$,
exhibits at short times a transient structure named {\it time domain resonance}. The maximum, $t_p$, of the {\it time domain resonance}
showed a plateau region for small $L$,
or more accurately  a shallow basin, 
followed by a linear dependence for larger $L$;
this behavior is reminiscent
of the Hartman effect, but the time of the plateau did not coincide 
with the phase-time estimate. It was found that for a broad range of parameters, $t_p$ in the basin may be written approximately as,
\begin{equation}
t_p^B=\frac{\hbar \pi}{\epsilon_1-E_0},
\label{tB}
\end{equation}
where $\epsilon_1$ and $E_0$, correspond to the energy of the first top-barrier resonance  and the incidence energy, respectively. Such a dependence had not been described before. On the other hand, along the linear regime, at larger values of $L$, $t_p$ is described by,
\begin{equation}
t_p^L=\frac{L}{v_1},
\label{tL}
\end{equation}
where $v_1=\hbar a_1/m$, with $a_1$ the real part of the first top-barrier pole $k_1=a_1-i b_1$. The above time scales are also
different from the BL-time, even though they 
may coincide with $t_p^L$  for a particular value of $L$.

We shall show in this paper that this basin 
dependence, and the corresponding 
time scale, is not only present at the transmission edge of the square barrier studied in \cite{gcv01}.
It may also be found, mutatis mutandis, in the sharp onset source model
examined in \cite{MB00} for 
small $x\kappa_0$. We also find the basin for the time of the forerunner 
versus position in the internal region of the square barrier, and for a step 
potential barrier. 
There are not resonances in the sharp onset source model, or for the step potential, 
so the time scale 
of the basin minimum is in these cases inversely proportional to $\kappa_0^2$,
namely to the difference between the ``potential level'' $V$ 
and the source
main energy $E_0$. This is to be contrasted 
with the dependence on $\kappa_0^{-1}$ of the traversal time $\tau$.

An important open question was to determine if the forerunner at 
small lengths in the evanescent region corresponds or not to tunneling
frequencies. We shall show by a simple time-frequency analysis
that the peak of the forerunner in this regime is composed predominantly
by under-the-barrier components, so that indeed a genuine tunneling
time scale different from phase or BL times has been found.

In section II the source with a sharp onset is discussed, and 
in section III we shall examine the internal region of the square barrier, as
well as the step barrier case. As the reader will soon discover, apart from certain 
peculiarities, all models show a small length region of the order of the penetration
length $\kappa_0^{-1}$ where the forerunner is dominated by tunneling components and 
arrives at a time proportional to $\kappa_0^{-2}$. 

\section{SOURCE WITH A SHARP ONSET}


\subsection{Formalism}

The time dependent solution of the 
Schr\"odinger equation for a potential $V(x)=V$ that
occupies the entire space, subject to a source 
boundary condition,  
%
%
%
\beq
\psi(0,t)=e^{-i\omega_0 t/\hbar}\Theta(t),
\label{initi}
\eeq

has the form, \cite{MB00} 
\beq\label{exact}
\Psi(x,t)=\frac{1}{2}e^{-itV/\hbar+ix^2/(4C^2t)}\left[
w(-u_0')+w(-u_0'')\right],
\eeq
where $w(z)=e^{-z^2} {\rm erfc} (-iz)$ and   
\beqa\label{us}
u_0'&=&\frac{1+i}{2^{1/2}}t^{1/2}C\kappa_0\left(-i-\frac{\tau}{t}\right),
\\
\nonumber
u_0''&=&\frac{1+i}{2^{1/2}}t^{1/2}C\kappa_0\left(i-\frac{\tau}{t}\right),
\\
\tau&=&x/v_{sc}=xm/\kappa_0 \hbar,
\\
C&=&(\hbar/2m)^{1/2}.
\eeqa
In the opaque limit ($x\kappa_0 \gg 1$), an excellent approximation
is given by adding up the pole and saddle terms, 
\beqa
\Psi(x,t)&\approx&\Psi_0+\Psi_s,
\\
\Psi_0&=&e^{-i\o0 t}e^{-\kappa_0 x}\Theta(t-\tau), 
\\
\Psi_s&=&\frac{1}{2i\pi^{1/2}}
e^{-\frac{itV}{\hbar}+\frac{ix^2}{4C^2t}}
\left(\frac{1}{u_0'}+\frac{1}{u_0''}\right).
\eeqa
For opaque conditions at fixed $x$ the saddle term is totally dominant in the
forerunner peak; at later times 
it fades away and the pole (tunneling) term
becomes dominant.
This allowed to identify $t_p$ from the peak of 
$|\Psi_s|^2$, and $t_{tr}$
as the time where the two contributions were of equal importance.  
The small $x\kappa_0$ regime did not permit the same approximation
treatment,  
so that $t_p$ could not be characterized by the formula found for
the opaque case, Eq. (\ref{tf}). It may be argued however, that the 
small $x\kappa_0$ case is in fact more interesting since it allows  
non-negligible signals (densities), whereas for large opacity 
the signal becomes exponentially small and very hard to detect.  
\subsection{Examples}

\begin{figure}
{\includegraphics[width=3.in]{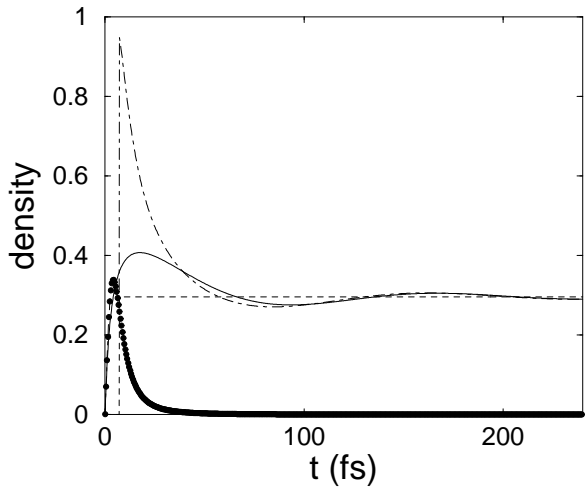}}
\caption[]{$|\Psi|^2$ (solid line), $|\Psi_s|^2$ (circles), 
$|\Psi_0|^2$ (dashed line), and $|\Psi_0+\Psi_s|^2$
(dotted-dashed line) 
versus $t$ for $E_0=0.907 V$, and $x=2.75$ nm. 
Here and in Figs. \ref{twfig}-\ref{etfig}, $V=0.3$ eV. In all figures
$m=0.067 m_e$, 
where $m_e$ is the electron's mass.}
\label{tdenfig}
\end{figure}

In Fig. \ref{tdenfig} we have depicted the density versus $t$ for a
fixed $x$ smaller than the penetration length $1/\kappa_0$
for the source-with-a-sharp-onset model. In all calculations the 
effective mass of the electron is taken as $m=0.067 m_e$.  
The forerunner is identified as a smooth broad bump with its peak at $t_p$.
The contributions of the saddle and pole terms are also drawn, as well 
as the contribution of their sum. The saddle term reproduces the density
only for very short times while the pole gives the large time 
behavior.    

\begin{figure}
{\includegraphics[width=3.in]{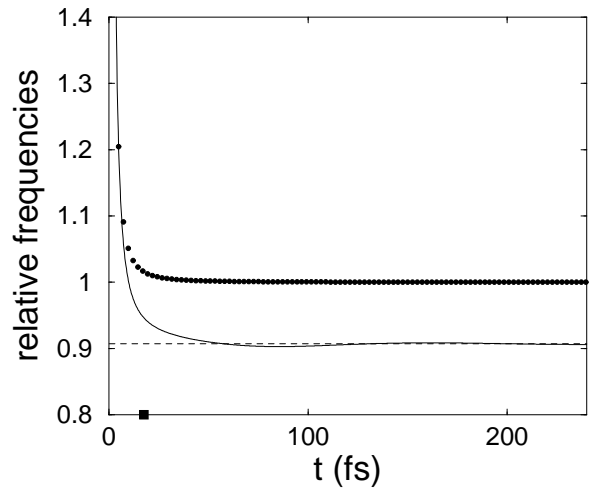}}
\caption[]{$\om_{av}/\om_V$ (solid line), $\om_0/\om_V$ (dashed line),
and $\om_s/\om_V$ (circles) versus 
$t$ for the same parameters of Fig. \ref{tdenfig}.
The square marks the value of $t_p$.}
\label{twfig}
\end{figure}
In Fig. \ref{twfig}
the average local frequency \cite{Cohen95,MB00},
$\om_{av}=-\textrm{Im}\left[(d\Psi/dt)/\Psi \right ]$, 
the frequency of the saddle 
$\om_s$, and $\om_0$ are shown, relative to $\om_V$,  versus $t$
for the same value of $x$.
Note that $\om_{av}$ tends to $\om_0$ for large $t$,
and to $\om_s$ for very short times, going from $\om_s$ to
$\om_0$ during the time span of the forerunner bump
(compare with the density in Fig. \ref{tdenfig});    
in particular $\om_{av}/\om_V<1$ at the peak $t_p$.   
\begin{figure}
{\includegraphics[width=3.in]{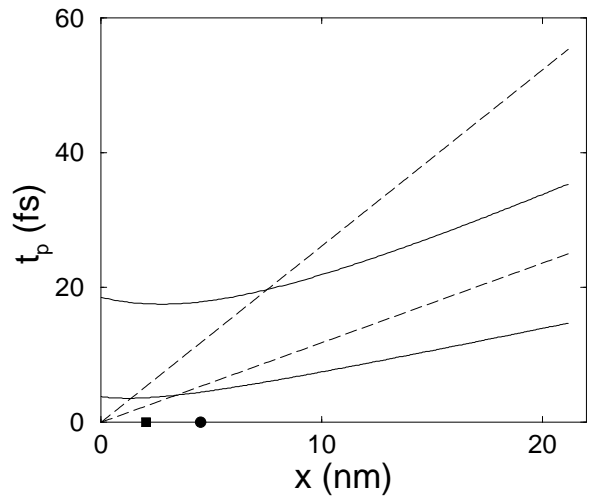}}
\caption[]{time of arrival of the forerunner's peak versus $x$ for 
$\omega_0/=0.907\omega_V$ (upper solid line),
and $\omega_0=0.544\omega_V$ (lower solid line). The upper and lower 
dashed lines are the corresponding BL-times.
The values of $1/\kappa_0$ are also 
shown with a square ($\omega_0=0.544\omega_V$)
and a circle ($\omega_0=0.907\omega_V$)}
\label{xtfig}
\end{figure}
Fig. \ref{xtfig} shows the dependence with $x$ of the 
time of arrival of the peak of the forerunner, 
$t_p$, for two different values of $\o0$.
The (unsharp) transition between the quasi-plateau region at lower 
$x$ and the linear regime for larger $x$ occurs around 
a few penetration lengths $1/\kappa_0$. Thus, a smaller $\omega_0$ implies 
a more reduced plateau. Also shown are the corresponding 
B\"uttiker-Landauer times for comparison.    
In the plateau region the forerunner's peak does not travel but arrives 
roughly simultaneously at all $x$ (in fact up to the basin minimum 
the forerunner arrives earlier at larger $x$!); beyond the plateau region 
the forerunner, whose peak is characterized in that case 
by Eq. (\ref{tf}), arrives later at larger $x$, and  becomes dominated by
frequencies ``above the barrier''.  
 
\begin{figure}
{\includegraphics[width=3.in]{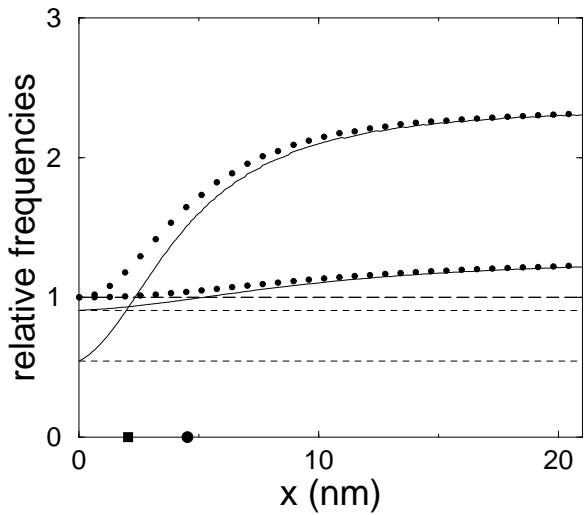}}
\caption[]{$\om_{av}/\om_V$ (solid line), $\om_0/\om_V$ (short dashed line), 
and $\om_s/\om_V$ (circles) for the same cases of Fig. \ref{xtfig}. 
Upper curves: $\omega_0=0.907\omega_V$; lower curves: 
$\omega_0=0.544\omega_V$.
The values of $\kappa_0^{-1}$ are indicated  as in Fig. \ref{xtfig}} 
\label{xwfig}
\end{figure}

\begin{figure}
{\includegraphics[width=3.in]{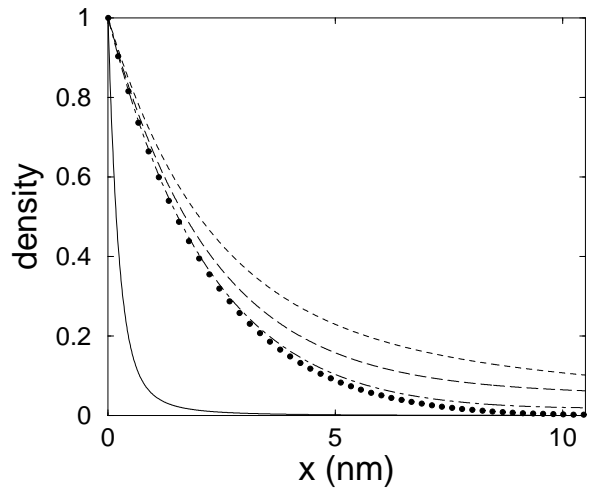}}
\caption[]{Density $|\Psi|^2$ versus $x$ for $\om_0/\om_V=0.907$
and $t=0.05,\, 24,\,48,\,72$ and $97$ ns (solid, short-dashed, 
long-dashed, dot-dashed, and circles respectively).}
\label{xdenfig}
\end{figure}

Fig. \ref{xwfig} shows the average frequency $\om_{av}$ calculated right at 
the forerunner's peak at the time $t_p(x)$ 
for each value of $x$, as well as $\om_s$,  
and $\om_0$, all relative to $\om_V$, 
for the same two values of $\om_0$ of the previous figure.     
The forerunner's peak frequency begins at $\om_0$ at $x\to 0$, then it 
grows smoothly and becomes non-tunneling ($\om_{av}/\om_V$ crosses $1$)
around $1/\kappa_0$. For larger $x$, 
$\om_{av}$ tends to the saddle frequency $\om_s$.    
Further insight into the nature of the tunneling forerunner 
is obtained by looking at a
series of snapshots of the wave density versus $x$ 
(instead of looking at density
versus time for a fixed location).
The plateau, or rather shallow basin,   
must be, according to Fig. (\ref{xtfig}), 
the result of a ``breathing'' transient mode of the evanescent wave,
which implies an essentially simultaneous growth of the wave for all 
$x$ up to a few penetration lengths, 
and until the maximum is achieved at $t_p$.
This is indeed confirmed in Fig \ref{xdenfig}, 
where several snapshots of the density are 
taken at five different times. In the final one the wave is  
already close to its asymptotic (large time) form.   
The linear dependence of the time of arrival of the peak 
with $\kappa_0^{-2}$ is shown in Fig. \ref{etfig}, which is drawn
by choosing the values of $V-E_0$ and $t_p$ corresponding to the minima of the 
shallow basins obtained for different values of $E_0$
(see Fig. {\ref{xtfig}).        
\begin{figure}
{\includegraphics[width=3.in]{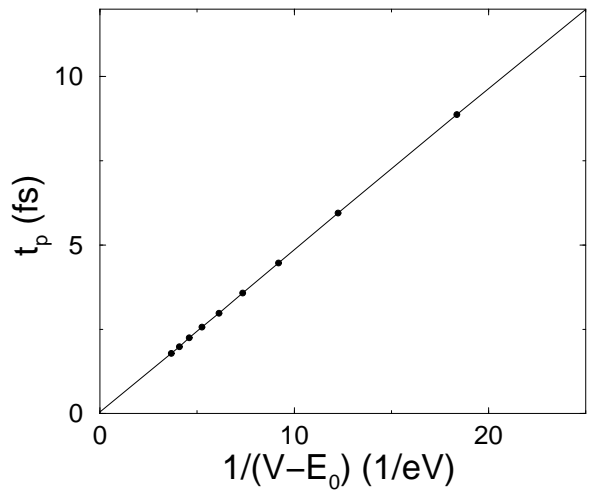}}
\caption[]{Time of arrival of the forerunner's peak $t_p$
versus $1/(V-E_0)$ at the position that minimizes $t_p$ for a given 
$E_0$, see Fig. \ref{xtfig}. As in all the figures of this section, 
$V=0.3$ eV.}
\label{etfig}
\end{figure}

\section{QUANTUM SHUTTER}

\subsection{Formalism}
Assume now an arbitrary potential 
$V(x)$ ($0\leq x\leq L$) that vanishes outside the internal region, and 
the initial condition,
\beq
\Psi \left( x,k;t=0\right) =\left\{ 
\begin{array}{cc}
e^{ikx}-e^{-ikx}, & \quad -\infty <x\leq 0, \\ 
0, & \quad x>0,
\end{array}
\right.  \label{obtur}
\eeq
which corresponds to a plane wave impinging on a perfectly reflecting shutter
placed at $x=0$, just at the left edge of the structure.
The time dependent process begins with the instantaneous opening of the shutter at $t=0$, 
enabling the incoming wave to interact with the potential at $t>0$. The exact solution
along the internal region ($0\leq x\leq L$) is given by \cite{gcr97},
\beq
\Psi ^{i}=\phi_{k}M(y_k)-\phi_{-k}M(y_{-k})-\sum\limits_{n=-\infty }^{\infty
}\rho_{n}M(y_{k_n}). 
\label{Psint}
\eeq
In the above expressions, the quantities $\phi_{\pm k}$ refer to the stationary solution,
and the factor $\rho_n=2iku_n(0)u_n(x)/(k^2-k_n^2)$ is given in terms of the
resonant eigenfunctions, $\{u_n(x)\}$, with complex eigenvalues \cite{gcr97}
$k_{n}=a_{n}-ib_{n}$ ($a_{n},b_{n}>0$).
\begin{figure}[!tbp]
\rotatebox{0}{\includegraphics[width=3.3in]{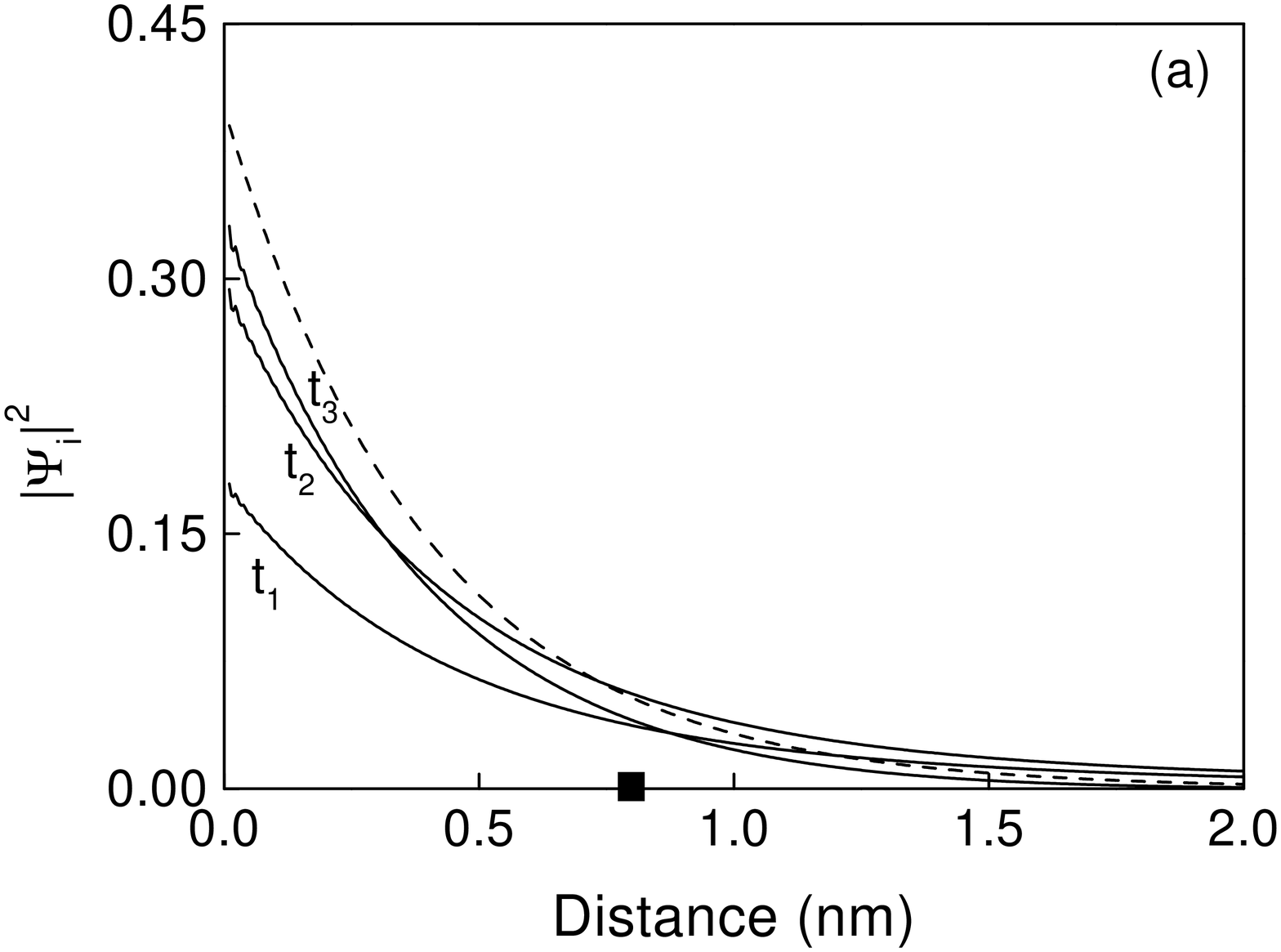}}
\rotatebox{0}{\includegraphics[width=3.3in]{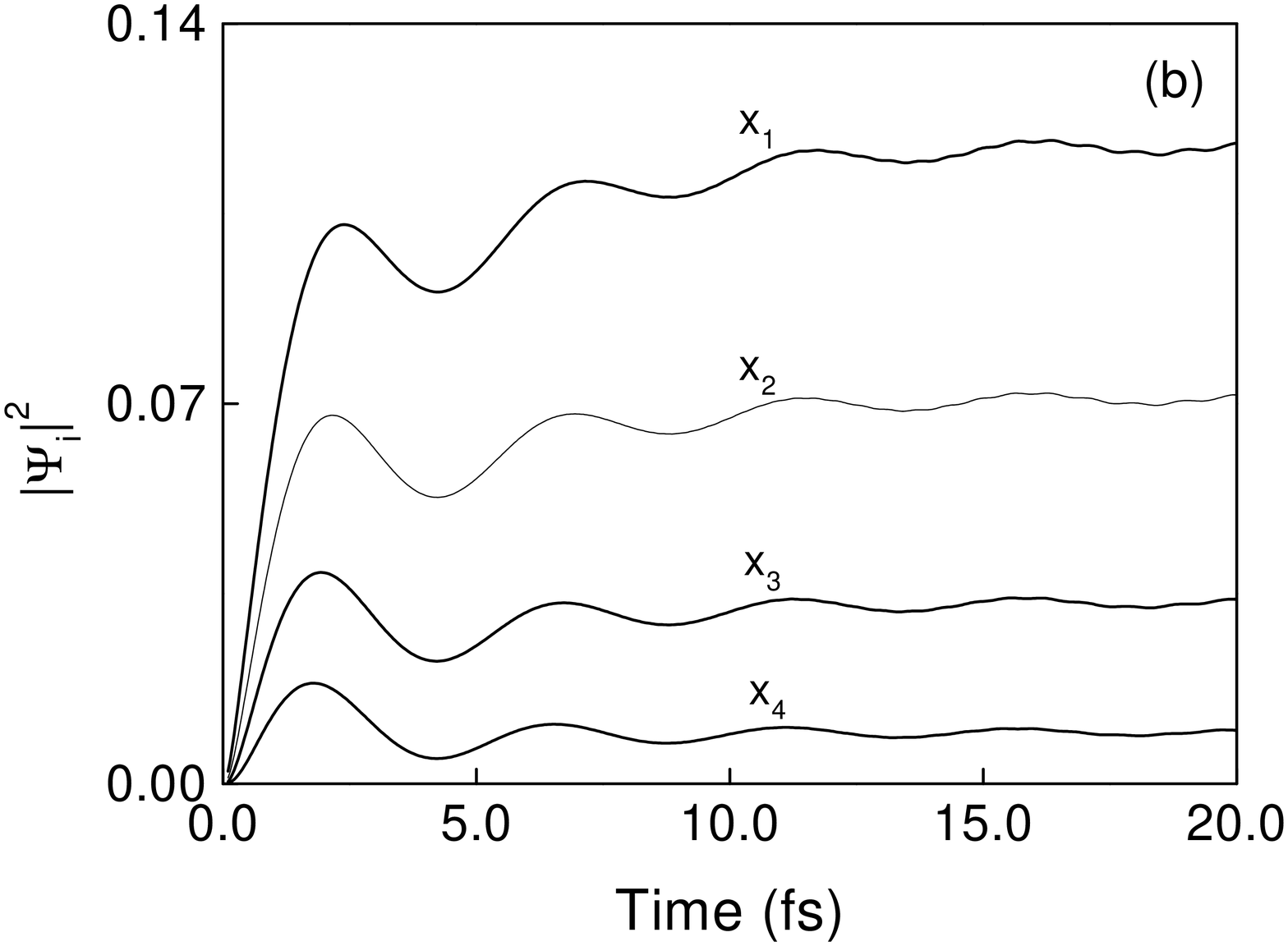}}
\caption{ (a) Snapshots of $|\Psi^{i}|^{2}$ (solid line)
taken at different values of time:
$t_1=1.0$ fs, $t_2=2.0$, and $t_3=4.0$ fs. The stationary solution $|\phi_k|^2$ is included
for comparison, and a full square indicates the value of $\kappa_0^{-1}$,
which is the penetration length of the stationary solution.
(b) Time evolution of $|\Psi^{i}|^{2}$ at different fixed values of the position $x$:
$x_1=0.5$ nm, $x_2=0.7$ nm, $x_3=1.0$ nm, and
$x_4=2.0$ nm. The solution exhibits a behavior similar to diffraction in time pattern.}
\label{smallex}
\end{figure}
The index $n$ runs over the complex poles $k_n$, distributed in the
third and fourth quadrants in the complex $k$-plane. Both the complex poles $\{k_{n}\}$
and the corresponding resonant eigenfunctions $\{u_n(x)\}$,
can be calculated using  a well established method,
as discussed elsewhere \cite{gcr97, gcv01}.
In the above equation the $M$ functions are defined as, 
\begin{equation}
M(y_{q})=\frac{1}{2}w(iy_{q}),
\label{mosh}
\end{equation}
where $y_{q}=-e^{-i\pi /4}[(\hbar/2m)
t]^{1/2}q$, $q=\pm k$, and $k_{\pm n}$.

\subsection{Examples}

In what follows we shall explore the features of the probability density along
the internal region of one-dimensional rectangular potential barriers of height
$V$ and thickness $L$, defined along the interval $0\leq x\leq L$.
We choose the following parameters: $V=1.0$ eV, $L=40.0$ nm,
and incidence energy $E_0=\hbar ^{2}k^{2}/2m=0.1$ eV. 
For this particular case the full barrier is opaque, 
$\alpha\equiv L[2mV]^{1/2}/\hbar=53.0$, but we are interested
in distances of the order of, or smaller 
than, the penetration length $\kappa_0^{-1}<<L$.  
\begin{figure}[!tbp]
\rotatebox{0}{\includegraphics[width=3.3in]{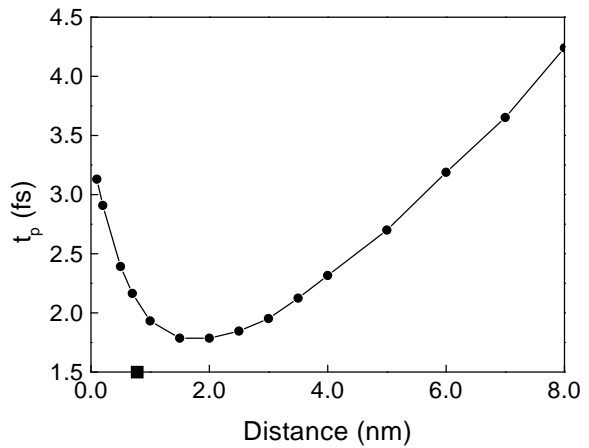}}
\caption{Exact calculation of $t_p$ (solid dots)
as a function of the position $x$. The parameters are the same
as in Fig. \ref{smallex}. A full square indicates the value
of $\kappa_0^{-1}$.}
\label{basin}
\end{figure}
In Fig. \ref{smallex} (a) we use Eq. (\ref{Psint})
to explore the birth of the probability density along
the internal region of the potential. Here
we plot $|\Psi^{i}|^{2}$ (solid line) as
a function of small values of the position $x$,
for increasing values of time.
Notice that as time increases, the probability density 
evanesces along the internal region
swinging around the stationary solution (dashed line). 

In Fig. \ref{smallex} (b) we illustrate the time dependence of the transient
``swinging mode'' exhibited in Fig. \ref{smallex} (a).
We exhibit the behavior of $|\Psi^i|^2$ (solid line) as
a function of time for several values of the position $x_i$ ($i=1,2,3,4$).  
The probability density grows monotonically until
it reaches its first maximum value,
and oscillates thereafter as it tends to the stationary situation given by 
$|\phi_k|^{2}$\cite{gcr97}.
Here we observe again a peculiar behavior in the time evolution
of  $|\Psi^{i}|^{2}$, presented in plot (b) of Fig. \ref{smallex}: the 
first maximum of the probability density appears earlier at 
larger values of $x$. In  Fig. \ref{basin}
we plot the exact calculation of $t_p$ (solids dots),
which corresponds to the peak value of the first maximum of $|\Psi^i|^2$,
measured at different values of the position, $x$. 

\begin{figure}[!tbp]
\rotatebox{0}{\includegraphics[width=3.3in]{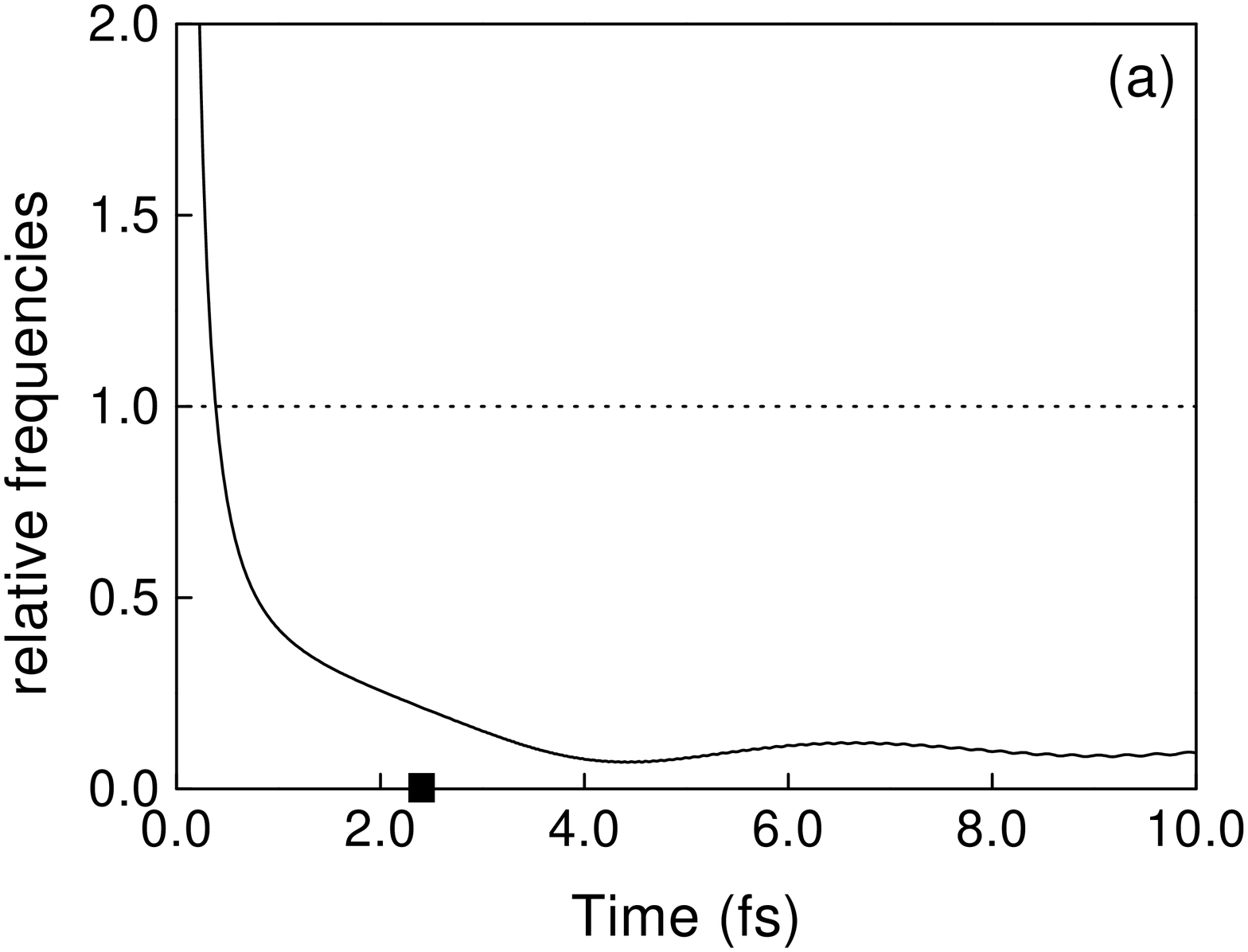}}
\rotatebox{0}{\includegraphics[width=3.3in]{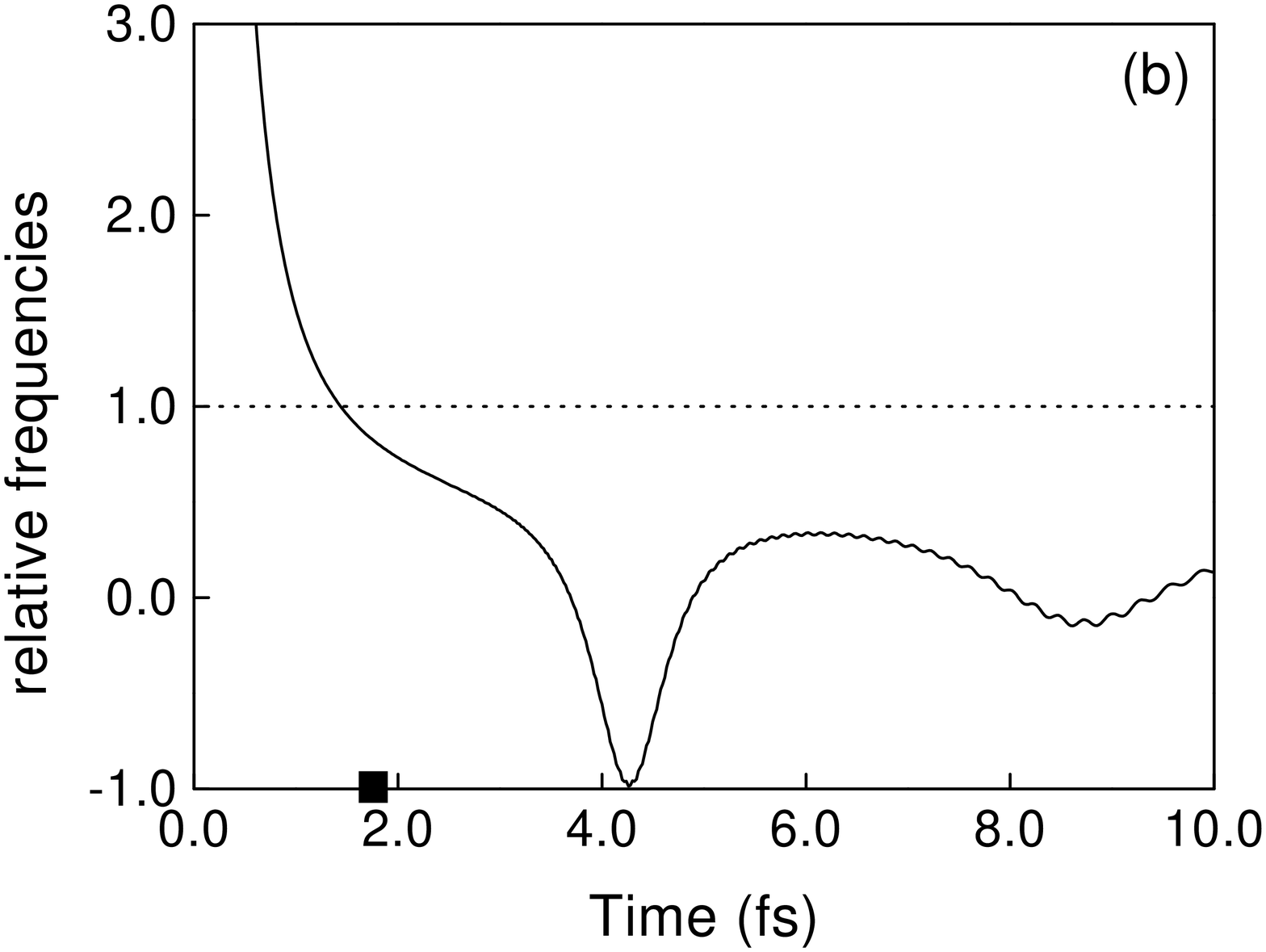}}
\caption{Plot of the relative local frequency $\omega_{av}/\omega_V$ 
(solid line) for the positions (a) $x_1=0.5$ nm, , and (b) $x_4=2.0$ nm, 
corresponding to the cases depicted in Fig. \ref{smallex} (b). The 
cutoff-frequency $\omega_V/\omega_V=1$ (dotted line) is included for 
comparison, and a full square indicates the position of $t_p$.}      
\label{espectro}
\end{figure}
In Fig. \ref{basin} we can clearly identify two regimes, as in the case
of Fig. \ref{xtfig}. The first of them corresponds to a basin,
where for small values of the position, $t_p$ decreases reaching a 
minimum value as $x$ increases.
This is in accordance with the observed behavior of the maximum
in Fig. \ref{smallex} (b). However, if $x$ is further increased, $t_p$
begins to grow as a function of the position.
Apparently this second regime corresponds to a situation where
$t_p$ increases linearly with $x$. The result depicted in Fig. 
\ref{basin} appears to be a replica of a similar behavior of the 
probability density at the barrier edge $x=L$,
recently reported in Ref. \cite{gcv01}. 

We have also analyzed the frequency content of $|\Psi^i|^2$ along 
the internal region. In Fig. \ref{espectro} (b) we plot 
$\omega_{av}/\omega_V$ along the relevant time interval for some 
of the values of the position, depicted in Fig. \ref{smallex} (b).
We can appreciate that  the probability density at this small 
values of the position $x$, is composed entirely by under-the-barrier
frequency components i.e $\omega_{av}/\omega_{V}<1$ in the vicinity 
of the maximum $t_{p}$. Notice that this result is similar to the 
one obtained in Fig. \ref{twfig}.

Finally, Fig. \ref{tfminima} shows the linear dependence of the 
time corresponding to the basin minimum with respect to $1/(V-E_0)$ 
in analogy to the analysis of the point source problem. Note, however, 
that in our case we could also have considered a plot with respect 
to $1/(\varepsilon_1-E_0)$ with a similar result since it turns out 
that for opaque barriers $\varepsilon_1 \approx V$.

\begin{figure}
\rotatebox{0}{\includegraphics[width=3.3in]{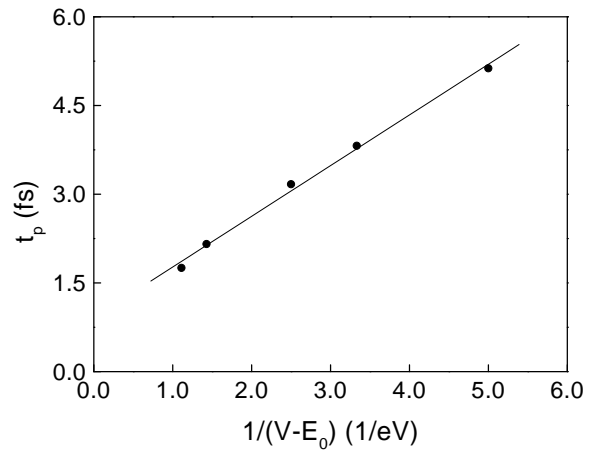}}
\caption[]{Maximum of the {\it time domain resonace}, $t_p$, as a function
of $1/(V-E_0)$. The parameters are the $V=1.0$ eV, and $L=40.0$ nm.}
\label{tfminima}
\end{figure}

\subsection{Potential step}

\begin{figure}
{\includegraphics[width=3.in]{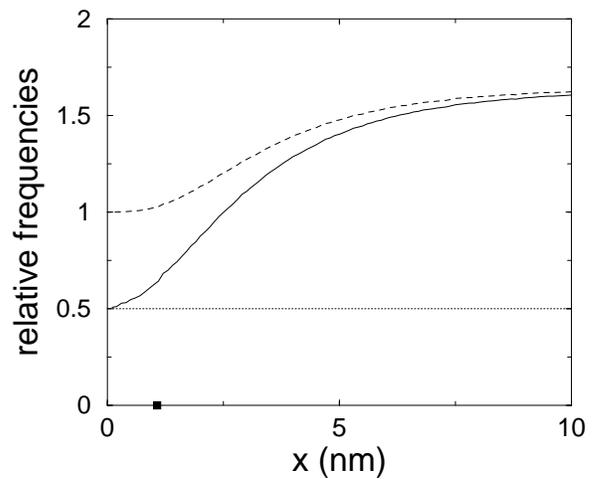}}
\caption[]{$\om_{av}/\om_V$ (solid line), $\om_0/\om_V$ (dotted line), 
and $\om_s/\om_V$ (dashed line) for $V=1$ eV and $E_0=0.5$ eV. 
The value of $\kappa_0^{-1}$ is indicated  with a square.} 
\label{frec}
\end{figure}

The potential step $V\Theta(x)$ requires a different formal 
treatment \cite{muga02} but very similar results are found.   
We have evaluated $\la x|\Psi(t)\ra$
by means of an integral over the energy eigenfunctions for the 
the initial state, 
\beq
\Psi \left(x,k;t=0\right) =\left\{ 
\begin{array}{cc}
e^{ikx}, & \quad -\infty <x\leq 0, \\ 
0, & \quad x>0.
\end{array}
\right.  
\eeq
In Fig. \ref{frec} the average frequency peak of the forerunner is shown versus 
$x$ for this system.
As in the source with a sharp onset case,
it goes from $\omega_0$ to $\omega_s$,  
crossing the threshold frequency at approximately $2/\kappa_0$.
Fig. \ref{tvE} shows also the linear dependence of
the time of the forerunner 
(calculated at its minimum value for a given energy,
{\it i.e.}, at the basin minimum)
versus $1/(V-E_0)$. 

\begin{figure}
{\includegraphics[width=3.in]{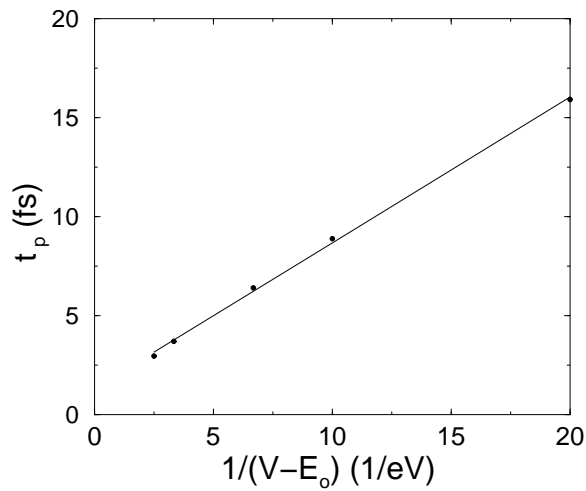}}
\caption[]{Time of arrival of the forerunner's peak $t_p$
versus $1/(V-E_0)$ at the position that minimizes $t_p$ for a given 
$E_0$. $V=1$ eV.}
\label{tvE}
\end{figure}

\section{Final comments}

By using different time-dependent models of quantum waves impinging
on a potential barrier, we have shown the existence of forerunners
dominated by under-the-barrier components. This transient structures
are indeed associated with a true tunneling process, occurring at 
distances smaller than the penetration length, $\kappa_0^{-1}$. 
We find that the time of arrival is proportional to $1/(V-E_0)$, 
where $E_0$ stands for the incidence energy, chosen below the
barrier height, $V$. This is in contrast to the dominance of above 
the barrier frequencies for larger distances, and to other well
known tunneling time scales such as the B\"uttiker-Landauer or
``phase'' times.

These results suggest a number of questions that need further 
investigation. For example, the proportionality constant observed
in the arrival times of the forerunner's peak,
depends on the peculiarities 
of each model, and a general theory for its specific value
should be found. Also, the Larmor time for spin rotation
defined by B\"uttiker is given for very thin barriers  
by $\hbar/V$ in the limit $k=0$ \cite{Buttiker83}. 
In spite of the very different ways in which the Larmor time
and the forerunner's peak time are obtained, these two quantities 
might be related.
Finally, we hope that the results presented may motivate an experimental 
search of tunneling forerunners and their corresponding time scale.

\begin{acknowledgments}
We are grateful to M. B\"uttiker for commenting on a preliminary version 
of the manuscript.  JGM and FD acknowledge support 
by Ministerio de Ciencia y Tecnolog\'\i a (BFM2000-0816-C03-03), 
UPV-EHU (00039.310-13507/2001), and the Basque Government (PI-1999-28).
GC and JV acknowledge financial support from Conacyt, M\'{e}xico, through 
Contract No. 431100-5-32082E, and GC that of DGAPA-UNAM under grant IN101301.
\end{acknowledgments}


\end{document}